\documentclass[12pt,aps,pra,groupedaddress,showpacs,amssymb,amsfonts,nofootinbib,tightenlines]{revtex4}
\usepackage{times,color}
\usepackage{ifthen,hyperref}

\definecolor{purple}{rgb}{0.625,0.125,0.9375}
\ifx\hypersetup\undefined\relax\else
\hypersetup{colorlinks=true,urlcolor=blue,linkcolor=purple,pageanchor=false}
\fi

\newcommand{\ignore}[1]{}
\newcommand{\todo}[1]{}
\newcommand{\ftodo}[1]{}
\newcommand{\mComment}[1]{}
\newcommand{\hComment}[1]{}
\newcommand{\gComment}[1]{}
\newcommand{\lComment}[1]{}

\ignore{
}
\ignore{ptexc cohere}

\usepackage{calc,ifthen}

\newlength{\elimdepthdim}
\newlength{\elimheightdim}
\newlength{\elimwidthdim}

\newlength{\strutdepthdim}
\newlength{\strutheightdim}
\newlength{\strutwidthdim}

\newcounter{herefignum}

\def\openone{\leavevmode\hbox{\small1\kern-3.8pt\normalsize1}}
\def\RR{{\rm I\kern-.2emR}}

\def\openone{\leavevmode\hbox{\small1\kern-3.8pt\normalsize1}}
\def\RR{{\rm I\kern-.2emR}}

\def\cg{{\cal G}}

\def\ch{{\cal H}}

\newcommand{\bitem}{\begin{itemize}}
\newcommand{\eitem}{\end{itemize}}
\newcommand{\benum}{\begin{enumerate}}
\newcommand{\eenum}{\end{enumerate}}
\newcommand{\beq}{\begin{equation}}
\newcommand{\eeq}{\end{equation}}
\newcommand{\beqa}{\begin{eqnarray}}
\newcommand{\eeqa}{\end{eqnarray}}
\newtheorem{definition}{Definition}
\newtheorem{proposition}{Proposition}
\newcommand{\eproof}{\end{proof}}
\newcommand{\bprop}{\begin{proposition}}

\newcommand{\bdef}{\begin{definition}}

\begin{document}

\title{ Random decoupling schemes for \\ quantum dynamical control 
and error suppression } 



\author{ Lorenza Viola }
\email{lorenza.viola@dartmouth.edu} 
\affiliation{Department of Physics and Astronomy, Dartmouth College,
6127 Wilder Laboratory, 
Hanover, New Hampshire 03755, USA }
\author{Emanuel Knill}
\email{knill@boulder.nist.gov}
\affiliation{ 
Mathematical and Computational Sciences Division,
National Institute of Standards and Technology, 
325 Broadway, Boulder, Colorado 80305, USA}



\date{Published 17 February 2005; PRL {\bf 94}, 060502 (2005)}

\begin{abstract}
We present a general control-theoretic framework for constructing and
analyzing random decoupling schemes, applicable to quantum dynamical
control of arbitrary finite-dimensional composite systems.  The basic
idea is to design the control propagator according to a random rather
than deterministic path on a group.  We characterize the performance
of random decoupling protocols, and identify control scenarios where
they can significantly weaken time scale requirements as compared to
cyclic counterparts.  Implications for reliable quantum computation
are discussed.
\end{abstract}



\maketitle


Dynamical decoupling offers a versatile control toolbox for quantum
dynamical engineering in both traditional settings like
high-resolution spectroscopy~\cite{HaeberlenBook} and quantum
information science~\cite{NielsenBook}.  Decoupling schemes operate by
subjecting the target system to a series of open-loop control
transformations, in such a way that the net evolution is coherently
modified to a desired one~\cite{Viola1999Dec}.  This combines
intrinsic design simplicity with the ability to avoid auxiliary memory
and measurement resources, while additionally enabling straightforward
integration with other passive~\cite{passive} or active~\cite{active}
quantum control techniques.  Applications of decoupling range from the
removal of undesired couplings in interacting quantum subsystems to
active decoherence control and symmetrization in open quantum
systems~\cite{Viola2002note}.  In particular, the use of decoupling
methods in conjunction with procedures for universal
control~\cite{Viola1999Control} provides a route to noise-suppressed
quantum computation based solely on unitary means. Remarkably, recent
advances support the potential for highly fault-tolerant control
architectures~\cite{Viola2003Euler,khodjasteh2004}.

So far, general formulations of the decoupling problem have been
restricted to {\em deterministic} control actions.  In the simplest,
so-called {\em bang-bang} setting, where the latter are
instantaneous rotations drawn from a group ${\cal G}$,
decoupling according to ${\cal G}$ is enforced by cycling the control
propagator through {\it all} group elements, translating into pulse
sequences with minimal length $T_c$ proportional to the size of ${\cal
G}$~\cite{Viola1999Dec}.  This suffers from two main
drawbacks. Because averaging requires traversing all of ${\cal G}$ in
a suitable sense, decoupling becomes very inefficient for large
groups, leading to unrealistically high control rates if the
interactions to be removed have a short correlation time $\tau_c$. 
Furthermore, it is not clear how to handle interactions which are
themselves fluctuating on time scales short compared to the averaging
period $T_c$.  These limitations severely constrain the practicality
of decoupling as a strategy for decoherence suppression in open
systems.

In this Letter, we propose to overcome the above limitations by
introducing a framework for {\em random dynamical decoupling}.
Physically, our approach takes inspiration from a naturally occurring
instance of a random decoupling process; that is, the self-averaging
of intermolecular interactions in gases and isotropic liquids due to
random translational and re-orientational
motions~\cite{HaeberlenBook}.  This intuition is cast in
control-theoretic language by requesting that the control propagator
follows a random but \emph{known} path on ${\cal G}$~\cite{sidenote}.
We show how random decoupling may be used to achieve a desired
coherent averaging and obtain a bound on worst-case performance. By
comparing to ordinary cyclic schemes, we find that in the presence of
rapidly fluctuating interactions and/or large control groups,
randomized design may prove superior.  From the point of view of
decoherence suppression, this not only establishes in general the
counterintuitive possibility to actively cancel noise using
randomness, but it also opens new prospects for significantly
mitigating time-scale requirements in a wide class of control systems.

{\it Random decoupling setting.} Let $S$ be a quantum system with
state space ${\cal H}_S$, dim(${\cal H}_S)=d<\infty$, evolving under 
an arbitrary, possibly time-dependent drift Hamiltonian $H_0(t)$.
Without loss of generality, we assume $H_0(t)$ to be traceless for 
all $t$.  We begin by constructing a random decoupling protocol for
effectively switching off the evolution due to $H_0(t)$, under the
assumption of perfect, unbounded control.  Let the available control
generate a discrete or continuous compact group ${\cal G}$, acting on
${\cal H}_S$ via a faithful, unitary, projective representation $\mu$,
$\mu(g)= \hat{g}$ for $g \in \cg$, $\mu({\cal G})= \hat{\cal G}$.  A
{\em random decoupler} uses control in ${\cal G}$ in two ways: first,
to establish a {\em logical} frame that is related to the physical one
(where $H_0(t)$ is specified) by an element of $\hat{\cg}$; second, to
rotate the system according to ${\cal G}$ randomly over time, by
following a random control path $U_c(t)$. Thus, both the past control
operations and the times at which they are applied are known, but the
{\em future} control path is random.

The essence of the random decoupling approach is to directly
depict the evolution of the system in the logical frame that
continuously follows the applied control. 
Let $\rho_S (t)= U(t) \rho_S (0) U^\dagger (t)$ describe the state of
the system in the physical frame, evolving under the action of both
the internal Hamiltonian and the controller, and let $\tilde{\rho}_S
(t) = {U}_c^\dagger (t) \rho_S (t) {U}_c (t)$ denote the corresponding
logical state, with $\tilde{\rho}_S (0)=\rho_S (0)$. Then the
evolution in the logical frame is fully specified by a propagator
($\hbar=1$)
$$ \tilde{U}(t)= U_c^\dagger (t) U(t) ={\cal T} \hspace*{-.5mm}
\exp\left \{ -i \hspace{-0.5mm} \int_0^t  du 
\tilde{H}(u) \right\} \:, $$
where $\tilde{H}(t)= U_c^\dagger(t) H_0(t) U_c(t)$.
Under the usual cyclicity assumption of deterministic
decoupling, $U_c(t+T_c)=U_c(t)$ for $T_c>0$, the physical and logical
frames stroboscopically coincide at times $t_N=N T_c$, $N\in {\mathbb
N}$. By contrast, random decoupling is intrinsically {\em acyclic}, 
and the control path almost never returns the system to the physical 
frame.  However, the available information about the past
control trajectory may be exploited to bring the state of the system 
back to the physical frame if desired.

{\it Error bounds.} To determine whether and how well random
decoupling succeeds at suppressing the dynamics due to $H_0(t)$ it is
necessary to compare the evolution under the propagator $\tilde{U}(T)$
over a time interval $T$ to the identity evolution, up to a global
phase. A natural measure is provided by the error probability for an
arbitrary pure initial state $P_S=|\psi\rangle\langle\psi|$ of
$S$. With respect to the random nature of the control path, the
{\sl a-priori} error probability can be expressed as an expectation
\begin{equation} 
\epsilon_T (P_S)= {\mathbb E} \left \{ \text{tr}_S 
\left( P_S^\perp \tilde{\rho}_S(T)\right) \right\} 
\mbox{}= {\mathbb E}\left \{ \text{tr}_S \left( P_S^\perp 
\tilde{U}(T) P_S\tilde{U}(T)^\dagger \right) \right\} \:,
\label{ep}
\end{equation}
where $P_S^\perp = \openone_S - |\psi\rangle\langle\psi|$ is the 
orthogonal complement of $P_S$ and ${\mathbb E}$ denotes ensemble 
average.  Then a {\em worst-case pure state error probability} 
may be defined as
\begin{equation}
\epsilon_T = \text{Max}_{P_S} \left \{ \epsilon_T(P_S) \right\}\:.
\label{et}
\end{equation}
A quantitative bound for $\epsilon_T$ is contained in the following:

\vspace*{1mm}

{\bf Theorem 1}. {\it Suppose that $(i)$ $\cg$ acts irreducibly on
$\ch_S$. $(ii)$ $U_c(t)$ is uniformly random for each $t$.
$(iii)$ For any $t,s >0$, $U_c(t)$ and $U_c(t+s)$ are independent for
$s > \Delta t$. $(iv)$ $|| H_0(t) ||_2$ is uniformly bounded in time
by $k >0$.  Then }
\begin{equation}
\epsilon_T = O \left({T \Delta t \, k^2}\right) 
\;\;\;\text{\it for}\;\;\; {T \Delta t \,k^2} \ll 1 \:.
\label{rbound}
\end{equation}

Here, $|| A ||_2=\text{Max}\,|\text {eig}(\sqrt{A^\dagger A} )|$, and
uniformly random is intended relative to the invariant Haar measure
$\nu_{\cg}$ on $\cg$, normalized such that $\nu_{\cg}
(\cg)=1$~\cite{footnote}.  While a rigorous proof of the above Theorem
is rather lengthy~\cite{epaps}, an outline of the underlying
strategy suffices for gaining physical insight.  The key step is to
realize that, in each of the integrals involved in the Dyson series
expansion of the time-ordered exponentials defining $\tilde{U}(T)$ and
$\tilde{U}^\dagger(T)$ in Eq. (\ref{ep}), the independence assumption
$(iii)$ effectively partitions the integration domain in two separate
regions: a volume $W_1(\Delta t)$, where none of the integration
variables is more than $\Delta t$ away from all the remaining ones;
and the complement $W_2(\Delta t)$, where this condition is violated
by at least one variable.  The expectation relative to such a variable
may be taken separately, leading, under the uniformity assumption
$(ii)$, to a contribution of the form
$${\mathbb E} \left \{ U^\dagger_c(t) H_0(t) U_c(t) \right \} =
\int_{\cg} \hspace*{-.4mm}d\nu_{\cg} \,\hat{g}^\dagger H_0(t)
\hat{g} \:, \;\;\; t \in (0, \Delta t) \:. $$ 
Notice, as the result of such an {\em ensemble} average, the 
appearance of the same dynamical $\cg$-symmetrization which, in 
standard deterministic schemes, is achieved through the {\em time} 
average over a cycle~\cite{Viola1999Dec,Zanardi-Sym}.  In particular, 
the irreducibility assumption $(i)$ implies maximal projection in the 
set of scalars. That is, for $X$ traceless,
\begin{equation}
\int_{\cg} \hspace*{-.4mm}d\nu_{\cg} \,\hat{g}^\dagger X 
\hat{g} = \frac{\text{tr}(X)}{d} \openone_S =0 \:. 
\label{gaveraging}
\end{equation}
As a consequence, all terms originating from $W_2(\Delta t)$ vanish,
and the desired upper bound to $\epsilon_T(P_S)$ may be determined by
estimating the volume of $W_1(\Delta t)$.  The irreducibility
assumption can of course be weakened.  As it turns out, the final
result (\ref{rbound}) for $\epsilon_T$ has a simple intuitive
explanation, which we defer until after we describe the corresponding
error bound for deterministic schemes.

From an implementation perspective, one may distinguish two main
scenarios, depending on whether the decoupler is specified by a
continuous or discrete control group $\cg$. In the former case, the
decoupling time scale $\Delta t$ is {\em defined} by the independence
requirement between $U_c(t)$ and $U_c(t+s)$, condition $(iii)$
entering as a design constraint. Note that bounded-strength controls
might suffice as long as $\Delta t$ is finite.  If ${\cal G}$ is
discrete, the required random walk of $U_c(t)$ may be enforced through
a sequence of equally spaced bang-bang pulses randomly drawn from
$\hat{\cg}$.  In this case, the independence requirement is
automatically satisfied by identifying $\Delta t$ with the separation
between consecutive kicks. Either way, it is important to stress that
random decoupling (unlike deterministic decoupling) places {\em no}
restriction on the temporal behavior of $H_0(t)$, only on its maximum
eigenvalue.

{\it Random decoherence suppression.} The above formalism can 
be extended to the suppression of noise effects arising from the 
coupling between the target system $S$ and an uncontrollable quantum
environment $E$. Let the total drift Hamiltonian be expressed in the
form $H_0(t)= \openone_S\otimes H_E + \sum_a J_a (t) \otimes B_a$,
where $H_E$ accounts for the (typically unknown) evolution of $E$ 
and the internal evolution of $S$ is included among the interaction
operators, with tr$(J_a (t))=0$ for all $t$. The action of the
decoupler is understood as $U_c(t)\otimes \openone_E$. Physically, 
it is meaningful to define a pure-state error probability that
depends only on the {\em reduced} state of $S$ in the logical frame.
That is, $\tilde{\rho}_S(T)$ in Eq. (\ref{ep}) is now calculated 
as $ \tilde{\rho}_S(T)=\text{tr}_E \{\tilde{U}(T) 
\tilde{\rho}_{SE}(0) \tilde{U}^\dagger(T)\}$, $\tilde{\rho}_{SE}(0)
={\rho}_{SE}(0)$ being the joint initial state and, as before, the 
logical propagator $\tilde{U}(t)$ describing the combined evolution
in a frame that explicitly removes the control field. By purifying
the environment, we can assume that $\rho_{SE}(0) =
P_S \otimes P_E$, both $P_S$ and $P_E$ being one-dimensional 
projectors.  The derivation of a bound for $\epsilon_T(P_S)$ may
be formally carried out following the same steps as in the uncoupled
case. It suffices to observe that Eq. (\ref{ep}) is 
equivalent to 
$$ \epsilon_T (P_S)={\mathbb E}\left \{ \text{tr}_{S,E} \left( 
P_S^\perp \otimes \openone_E \tilde{U'}(T) 
P_S \otimes P_E \tilde{U'}(T)^\dagger \right) \right\} \:, $$
with the propagator
$$\tilde{U'}(t)= U_E^\dagger (t) U_c^\dagger (t) U(t) = {\cal T}
\hspace*{-.5mm}\exp\left \{ -i\hspace{-0.5mm} \int_0^t du
\tilde{H'}(u) \right\} \: $$ 
describing the evolution in a frame where {\em both} the applied 
control and the environment dynamics $U_E(t)=\exp(-i H_E t)$ are 
explicitly removed, and 
$\tilde{H'}(t)=\sum_a U_c^\dagger (t) J_a(t) U_c(t) \otimes B_a$.  
We thus have the following:

\vspace*{1mm}

{\bf Theorem 2}.  {\it Let $\cg$ act irreducibly on $\ch_S$ and 
satisfy the same uniformity and independence assumptions as in 
Theorem 1. If $|| \sum_a J_a (t) \otimes B_a ||_2$ is uniformly 
bounded in time by $\lambda >0$, then }
\begin{equation}
\epsilon_T = O \left({T \Delta t \, \lambda^2}\right) 
\;\;\;\text{\it for}\;\;\; {T \Delta t \, \lambda^2} \ll 1 \:.
\label{rdbound}
\end{equation}

Formally, $\lambda$ is a measure of the overall {\em noise 
strength} as defined in the context of quantum error correction
theory~\cite{Knill-NS}. As pointed out in this reference, caution 
is required in treating infinite-dimensional environments.
Physically, $1/\lambda=\tau_c$ is of the order of the shortest 
correlation time scale present in the interaction to be removed.  
While the latter provides the relevant time scale to consider in 
the absence of additional information about the environment's 
initial state, power spectrum, and internal dynamics, such 
properties may critically impact the decoupling performance in 
actual applications~\cite{1overf}. Thus, {\em lower} error bounds 
tend to be fairly example specific.

According to the above Theorems, $\epsilon_T$ can in principle be 
made arbitrarily small by appropriate control design, implying the
possibility to {\em arbitrarily suppress on average} the unwanted
evolution in the logical frame.  This is especially surprising for
decoherence suppression considering that, in the physical frame, the
applied random field appears to be in general a source of decoherence.
It is worth noting that the possibility to exploit randomization
was considered earlier for specific decoupling problems.  Preservation
of coherence of a lossy radiation mode via the random modulation of a
system parameter was established in~\cite{Mancini2002}.  More
recently, a randomized refocusing algorithm was proposed 
in~\cite{Bremner2004} in the context of efficient simulation of
quantum computation starting from few-body Hamiltonians on $n$ qubits.
While revisiting such specific situations in the light of the
present analysis is interesting in itself, our main goal in what 
follows is to continue developing a model-independent formulation
of random decoupling in general control-theoretic terms.

{\it Comparison with cyclic decoupling.}  In order to assess 
the performance and usefulness of random decoupling schemes, 
a comparative error bound for deterministic decoupling is needed. 
We focus on the standard situation where the drift Hamiltonian 
$H_0$ is time-independent, and decoupling is accomplished 
by cyclic averaging over a finite group of order $|\cg|>1$.  
Apart from the redundant ensemble expectation, 
Eqs. (\ref{ep})-(\ref{et}) still define a valid worst-case pure 
state error probability. The deterministic counterpart to Theorem 1 
is then the following:


{\bf Theorem 3}. {\it Suppose that $(i)$ $\cg$ acts irreducibly on
$\ch_S$. $(ii)$ $U_c(t)$ is assigned according to a cyclic path over
${\cg}$, with $U_c(t)=\hat{g}_j$ for $t\in [j\Delta t, j+1\Delta t)$,
$j=0,\ldots,{|{\cal G}|-1}$, $\Delta t>0$, and $T_c= |\cg|\Delta
t$. $(iii)$ $|| H_0 ||_2$ is bounded by $k >0$, with $k T_c <1$. 
Then }
\begin{equation}
\epsilon_T = O \left( \left({T T_c\, k^2}\right)^2 \right) 
\;\;\;\text{\it for}\;\;\; {T T_c \,k^2} \ll 1 -{k T_c} \:.
\label{dbound}
\end{equation}

The proof follows from a direct evaluation of the logical 
propagator $\tilde{U}(T)$ using average Hamiltonian 
theory~\cite{HaeberlenBook},
$$ \tilde{U}(T) =e^{-i \overline{H}T} \:, \;\;\;
\overline{H}=\sum_{\ell=0}^\infty \overline{H}^{(\ell)}\:, $$ 
where $\overline{H}$ is computed from the Magnus expansion under the
averaging and convergence conditions, $\overline{H}^{(0)}=0$ and $k
T_c <1$, respectively~\cite{epaps}.  We now provide an intuitive
justification to the error bounds we found.

Write $R= T \Delta t k^2 = (k \Delta t)^2 (T/\Delta t)$. For 
the random method, each control step can accumulate an error 
amplitude of up to $k \Delta t$. Randomizing the decoupler has 
the net effect that the amplitudes add up probabilistically. 
Therefore, over an evolution time $T$, the total error 
probability is bounded by the number $T/\Delta t$ of such 
intervals, times the error probability $(k \Delta t)^2$ 
of each step. Notice that the bound of Theorem 1 is indeed  
$\epsilon_T^R=O(R)$.  

For the cyclic method using $|\cg|$ steps of duration 
$\Delta t$ in each cycle, the dominant errors are due to 
$\overline{H}^{(1)}$. That is, they arise from non-commuting
contributions associated with pairs of intervals in a cycle.  
Thus, for each cycle the error amplitude is bounded by 
$|\cg|^2 (k \Delta t )^2$, and a total time $T$ contains 
$T/(|\cg| \Delta t)$ such cycles. If, as assumed, each cycle 
is identical and the interaction is constant, the total 
error amplitude is bounded by the sum, yielding $|\cg| R$. 
By squaring and using that $|\cg| \Delta t =T_c$, the bound 
of Theorem 3 emerges, $\epsilon_T^D=O(|\cg|^2 R^2)$. 

The above analysis shows that the worst-case errors of
the two procedures compare as follows: 
$$ \epsilon^R_T=O(R)  \;\;\;\text{vs}\;\;\;
   \epsilon^D_T=O\left( (|\cg|^2 R) R\right) \:, $$ 
the quantity $|\cg|^2 R$ becoming a relevant figure of merit for 
performance.  Thus, cyclic decoupling tends to perform better if 
any time dependence or fluctuations in the interactions to be 
removed have time scale {\em longer} than $T_c=|\cg|\Delta t$ 
and, in addition, $|\cg|^2 R \ll 1$. Superior performance 
of random decoupling is expected instead in situations where 
the effective correlation or fluctuations have time scales large 
compared to $\Delta t$ but {\em short} compared to $|\cg|\Delta t$;
or, alternatively, $|\cg|^2 R \gg 1$. 

{\it Generalizations and applications.} The above results lend
themselves to a number of generalizations. The extension to reducible
group actions (hence selective decoupling) is conceptually
straightforward. Procedures for universal decoupled control may be
designed similarly to~\cite{Viola1999Control}, by randomly modulating
the applied control Hamiltonians to compensate for the decoupler
action if necessary.  This paves the way to schemes for randomly
controlled noise-suppressed universal quantum computation. In
addition, one may envisage a variety of hybrid control schemes where
deterministic and random operations are simultaneously exploited.  At
least two options are worth considering.  First, one may randomize the
decouplers.  If multiple decouplers are available to effect a desired
averaging, which one to apply may be picked at random at every
cycle. Or, with a single decoupler, one may randomize the cycles, by
randomly choosing which path to follow to traverse $\cg$.  While a
clever concatenation of deterministic and random protocols could merge
advantageous features from both methods, quantitative error estimates
as well as studies of the {\em typical} performance in specific
situations will be reported elsewhere.

We anticipate that randomization might offer substantial benefits
whenever a large number of control time-slots is involved.  An extreme
example is maximal decoupling in $n$ arbitrarily coupled qubits,
$d=2^n$.  Deterministic group-based schemes require averaging over 
the Pauli error basis $\{\openone, \sigma_x, \sigma_y,
\sigma_z\}^{\otimes\,n}$, with $|\cg|=
d^2=4^n$~\cite{Viola1999Dec}. For fixed control parameters $T, \Delta
t$ such that $R \ll 1$, the condition $|\cg|^2 R \ll 1$ becomes
exponentially harder to meet as $n$ increases.  Equivalently, for a
fixed tolerable error $\epsilon_T$, an interval $\Delta t$ that
shrinks exponentially with $n$ is needed to compensate $|\cg|^2$ in
this case. A randomized implementation of Pauli decoupling is indeed
at the heart of the simulation algorithm mentioned
above~\cite{Bremner2004}.  In addition, the recently proposed
Pauli-Random-Error-Correction method for coherent
errors~\cite{Kern2004} may also be understood as an ingenious
application of the present control framework, random Pauli rotations
being repeatedly applied to average {\it static} imperfections, and
permutations of the original logic gates ensuring the intended
decoupled control. While cyclic schemes with quadratic
complexity~\cite{Roetteler} are known for bilinearly coupled qubits as
assumed in~\cite{Kern2004}, randomized schemes may still be attractive 
for large $n$ and/or time-varying couplings. In the same spirit, the
cancellation of rapidly fluctuating {\it dynamical} imperfections
reported in~\cite{facchi2004Dyn} may be suggestively reinterpreted as
a random self-decoupling effect.  Lastly, efficiency improvements are
to be expected from decoupling according to the symmetric group ${\cal
S}_n$ acting on $n$ qubits, which otherwise involves factorial
overheads, and is relevant to the synthesis of collective
noise~\cite{Collective}.

{\it Conclusion.} We introduced an approach to dynamical decoupling
that relies on random control design. Beside being interesting {\em
per se} as a largely unexplored setting for coherent and error control,
random dynamical decoupling carries the potential for faster
convergence and relaxed timing constraints compared to deterministic
counterparts in relevant situations.  While additional work is needed
to expand the present analysis, we believe that our results add to the
significance of decoupling methods as a control-theoretic tool and
allow a step forward toward making them a practical error control
strategy in quantum information science.

L.~V. acknowledges support from LANL during the early stages of this
work. E.~K. was supported by the US NSA. We thank Howard Barnum for
discussions. Contributions to this work by NIST, an agency of the US
government, are not subject to copyright laws.

\vspace*{-3mm}


\begin{thebibliography}{21}
\bibitem[{sid()}]{sidenote}
\bibinfo{note}{Knowledge of the path may be ignored if only operators commuting
  with ${\cal G}$ are used, as is usually the case in the context of
  self-averaging of intermolecular interactions.}
\end{thebibliography}

\vspace*{5mm}

\begin{appendix}

{{\bf Appendix }}\hrulefill

\vspace{1mm}

In this Appendix, we supply the proofs of Theorem 1 and 3. Theorem
2's proof can be obtained from that of Theorem 1 by calculating
the relevant worst-case error probability in the appropriate doubly-rotating
frame.

{\it Proof of Theorem 1}.
Let $P_S$ be an arbitrary pure state of $S$. The logical propagator
$\tilde{U}(T)$ may be expressed as follows:
\begin{equation}
\tilde{U}(T)= \sum_{n=0}^\infty I_n(T)\:,
\end{equation}
where 
\begin{equation}
I_n(T) =(-i)^n \int_{0\leq u_1\leq\ldots\leq u_n\leq T} du_n\ldots du_1 
   \tilde{H}({u_n}) \ldots \tilde{H}({u_1})\:,
\end{equation}
and similarly for $\tilde{U}(T)^\dagger$. Thus, we need to calculate 
\begin{eqnarray}
\epsilon_T(P_S) = {\mathbb E}\left \{ \text{tr}_S \left( 
   \sum_{n,m=0}^\infty P_S^\perp I_n (T) P_S I_m(T)^\dagger \right)
\right\} =
   \sum_{n,m=0}^\infty {\mathbb E}\left \{ \text{tr}_S \left( 
 P_S^\perp I_n (T) P_S I_m(T)^\dagger \right) \right\}  \:.
\end{eqnarray}
The contributions with $n=0$ or $m=0$ vanish because of $P_S^\perp$ and 
$P_S$ cancel each other upon exploiting the cyclicity of the trace.
By noticing that $\epsilon_T(P_S)\geq0$ hence $\epsilon_T(P_S)=
|\epsilon_T(P_S)|$, 
\begin{equation}
\epsilon_T(P_S) \leq \sum_{n,m \geq 1} 
\left| {\mathbb E}\left \{ \text{tr}_S \left( 
 P_S^\perp I_n (T) P_S I_m(T)^\dagger \right) \right\}  \right|\:.
\end{equation}
Under the assumption of sufficiently smooth behavior, the expectation 
may be moved under the integral. Fix a pair of integers $n,m \geq 1$,
then the relevant contribution is 
\begin{equation}
\int_{0\leq \ldots\leq u_n\leq T; \, 
0\leq \ldots \leq t_m\leq T} du_1\ldots du_n  dt_1\ldots dt_m\,
{\mathbb E}\left \{ P_S^\perp \tilde{H}(u_n) \ldots \tilde{H}(u_1)
P_S\tilde{H}(t_1)\ldots \tilde{H}(t_m)\right\}\:. 
\label{nm}
\end{equation}
Let $W_1^{(n,m)}(\Delta t)$ denote the set of points 
$(u_1,\ldots,u_n,t_1,\ldots, t_m)$ satisfying that $u_\ell$ and $t_\ell$ 
are each time-ordered and no $u_\ell$ or $t_\ell$ is further away than
$\Delta t$ from the rest, and let $W_2^{(n,m)}(\Delta t)$ denote the 
remaining integration volume in Eq. (\ref{nm}). Because, within 
$W_2^{(n,m)}(\Delta t)$, at least one of the integrating variables is
more than $\Delta t$ away from all the other variables, the 
independence assumption $(iii)$ allows the expectation relative to 
such a variable to be taken separately.  By the uniformity assumption
on $U_c(t)$ for all $t$, and by the tracelessness assumption on
$H_0(t)$ for all $t$, such an expectation vanishes. Therefore, 
$W_1^{(n,m)}(\Delta t)$ is the only subset of points contributing to 
the expectation in Eq. (\ref{nm}). Let $dw^{(n,m)}$ denote the 
corresponding integration measure. Then
\begin{eqnarray}
\epsilon_T(P_S) &\leq& \sum_{n,m \geq 1} \int_{W_1^{(n,m)}} 
dw^{(n,m)}\, \left| {\mathbb E}\left \{ \text{tr}_S \left( 
 P_S^\perp \tilde{H}(u_n) \ldots P_S \ldots \tilde{H}(t_m) \right) \right\}  
\right| \nonumber \\
 & \leq & \sum_{n,m \geq 1} \int_{W_1^{(n,m)}} 
dw^{(n,m)}\,  {\mathbb E}\left \{ \left| \text{tr}_S \left( 
 P_S^\perp \tilde{H}(u_n) \ldots P_S \ldots \tilde{H}(t_m) \right)\right|\right\}  
\:,
\end{eqnarray}
where in the second step Jensen's inequality has been used. By noticing
that the argument of the trace is a rank-1 operator, one can simplify
\begin{eqnarray} 
\epsilon_T(P_S) &\leq&
 \sum_{n,m \geq 1} \int_{W_1^{(n,m)}} 
dw^{(n,m)}\,  {\mathbb E}\left \{ || 
 P_S^\perp \tilde{H}(u_n) \ldots P_S \ldots \tilde{H}(t_m) ||_2 \right\}
\nonumber \\
& \leq & 
\sum_{n,m \geq 1} \int_{W_1^{(n,m)}} 
dw^{(n,m)}\,  {\mathbb E}\left \{ || P_S^\perp ||_2 \, || P_S ||_2 \,
k^{n+m} \right\} \nonumber \\
&\leq&  \sum_{n,m \geq 1} {\text {Vol}}({W_1^{(n,m)}}) k^{n+m} \:, 
\end{eqnarray}
where the uniform bound $k$ for $H_0(t)$ has been used, and 
{\text Vol}$({W_1^{(n,m)}})$ is the volume of ${W_1^{(n,m)}}$. Note
that the dependence upon the initial state $P_S$ has disappeared at
this point.

The above volume may be estimated through the following combinatorial
argument.  First, notice that given the two ordered lists
$0\leq u_1 \leq\ldots\leq u_n\leq T$, 
$0\leq t_1 \leq\ldots\leq t_m\leq T$, there are ${n+m\choose m}$
different merged orderings. Fix a particular one. Then each element 
needs to be either within $\Delta t$ of the next one or of the previous
one. Make a choice for the odd-numbered elements, the first element
being labeled $1$.  There are at most $2^{\lceil (n+m)/2\rceil}$ such
choices.  For each of them the contribution to the volume may be 
bounded by ordering the even-numbered elements, then by inserting the 
odd ones, ignoring the ordering constraint now.  Finally, 
\begin{eqnarray}
{\text {Vol}}({W_1^{(n,m)}}) &\leq &
{n+m\choose m} 2^{\lceil (n+m)/2\rceil} \frac{T^{\lfloor(n+m)/2\rfloor}
\Delta t^{\lceil (n+m)/2 \rceil}}{(\lfloor (n+m)/2\rfloor)!} \nonumber \\
&\leq & 2^{\lceil (n+m)/2\rceil} T^{\lfloor (n+m)/2\rfloor} 
(2\Delta t)^{\lceil (n+m)/2\rceil}\:,
\end{eqnarray}
where the inequalities ${n+m\choose m}\leq 2^{n+m-1}$ (for $n+m\geq 2$),
and $\lfloor (n+m)/2\rfloor ! \geq 2^{\lfloor (n+m)/2 \rfloor -1}$ have 
been exploited.

The last step is to sum over $n,m$:
\begin{equation}
\text{Max}_{P_S} \{\epsilon_T(P_S) \} = \epsilon_T \leq 
\sum_{n,m =1}^\infty
 2^{\lceil (n+m)/2\rceil} T^{\lfloor (n+m)/2\rfloor} 
(2\Delta t)^{\lceil (n+m)/2\rceil} k^{n+m} \:. 
\label{fin}
\end{equation}
This may be done by considering separately the four partial sums 
where both $n$ and $m$ have the same (even or odd) parity, or they
have opposite (even-odd or odd-even) parity, respectively, and by
evaluating the $\lfloor\, \rfloor$, $\lceil \, \rceil$ in Eq. 
(\ref{fin}) accordingly. Straightforward calculations yield
\begin{equation}
\epsilon_T \leq (4T\Delta t k^2) \frac{1+ 8 \Delta t k + 4T\Delta t k^2 }
{(1-4T\Delta t k^2)^2} = O(T \Delta t k^2) 
\end{equation}
for values of $T \Delta t k^2 \ll 1$, as quoted in 
Eq. (\ref{rbound}).\hfill$\Box$

\vspace{1mm}

{\it Proof of Theorem 3}. The logical propagator $\tilde{U}(T)$
may be expressed in terms of the average Hamiltonian $\overline{H}$
as 
\begin{equation}
\tilde{U}(T)= e^{-i \overline{H} T} = \sum_{n=0}^\infty 
\frac{1}{n!} (-i \overline{H} T)^n \:,
\end{equation}
and similarly for $\tilde{U}(T)^\dagger$. Then the desired pure-state
error probability is bounded by
\begin{equation}
\epsilon_T(P_S) \leq   \sum_{n,m=1}^\infty \frac{1}{n!} \frac{1}{m!}
\left| \text{tr}_S \left( 
P_S^\perp ( -i \overline{H} T )^n  P_S ( +i \overline{H} T )^m \right)
\right| \:,
\end{equation}
where the orthogonality of the projectors $P_S, P_S^\perp$ has been
used to remove the terms with $n=0$ and/or $m=0$. By observing that
the argument of the trace is a rank-1 operator this gives
\begin{equation}
\epsilon_T(P_S) \leq  \sum_{n,m=1}^\infty \frac{1}{n!} \frac{1}{m!}
|| P_S^\perp||_2 \,||P||_2\, || \overline{H} T ||_2^{n+m}  \leq
\left( 1-e^{|| \overline{H}||_2 T} \right)^2 \:.
\end{equation}
Assuming that first-order averaging and convergence conditions for the 
Magnus series are fulfilled, one has $\overline{H}^{(0)}=0$ and $k T_c 
<1$, implying 
$$  || \overline{H}||_2 \leq \sum_{\ell=1}^\infty k (k T_c)^\ell 
= \frac{k^2 T_c}{1-k T_c}\:. $$
In the limit of sufficiently short time, $k^2 T T_c < 1-k T_c$, 
$ || \overline{H}||_2 T <1$ and by using the inequality 
$|1-e^x| \leq |x|/|1-x|$ for $x <1$ one obtains
\begin{equation}
\text{Max}_{P_S} \{\epsilon_T(P_S) \} = \epsilon_T \leq 
\frac{ ( || \overline{H}||_2 T)^2}{(1-|| \overline{H}||_2 T)^2}
=O\Big( (|| \overline{H}||_2 T)^2 \Big) = O\Big( (T T_c k^2)^2 \Big)
\end{equation}
for $T T_c k^2 \ll 1-kT_c$, as quoted in Eq. (\ref{dbound}).\hfill$\Box$

\end{appendix}

\end{document}